\documentclass[aps,prl,reprint,groupedaddress]{revtex4-1}
\usepackage{graphicx}

\usepackage{color}
\usepackage{amsmath}
\definecolor{Blue}{rgb}{0.00, 0.00, 1.00}
\definecolor{Red}{rgb}{1.00, 0.00, 0.00}

\begin{document}

\title{Ising model with stochastic resetting}

\author{Matteo Magoni$^1$$^{,}$$^2$, Satya N. Majumdar$^1$, and Gr\'egory Schehr$^1$}

\affiliation{$^1$Universit\'e Paris-Sud, CNRS, LPTMS, UMR 8626, 91405 Orsay, France \\
$^2$Department of Applied Science and Technology, Politecnico di Torino, Corso Duca degli Abruzzi 24, 10129 Torino, Italy}

\date{\today}

\begin{abstract}
We study the stationary properties of the Ising model that, while evolving towards its equilibrium state at temperature $T$ according to the Glauber dynamics, is stochastically reset to its fixed initial configuration with magnetisation $m_0$ at a constant rate $r$. Resetting 
breaks detailed balance and drives the system to a non-equilibrium stationary state where the magnetisation acquires a nontrivial distribution, leading to a rich phase diagram in the $(T,r)$ plane. We establish these results exactly in one-dimension and present scaling arguments supported by numerical simulations in two-dimensions. We show that resetting gives rise to a novel ``pseudo-ferro'' phase in the $(T,r)$ plane for $r > r^*(T)$ and $T>T_c$ where $r^*(T)$ is a crossover line separating the pseudo-ferro phase from a paramagnetic phase. This pseudo-ferro phase is characterised by a non-zero typical magnetisation and a vanishing gap near $m=0$ of the magnetisation distribution.   
\end{abstract}






\maketitle

Stochastic resetting has seen enormous activities during the last few years \cite{reset_review}, notably in the context
of search processes which are ubiquitous in nature \cite{Benichou2011}. The simple intuition behind resetting is
as follows. If one is searching for a target via a stochastic process such as simple diffusion,  
it may take a long time due to trajectories that run off from the target. It is then advantageous to restart the
search process with a certain resetting rate $r$ from the same initial condition~\cite{SMdiff,SMopt}. The idea is that one may 
explore new pathways leading to the target, thereby reducing the search time. Recently, a large number of studies have shown that there is
an optimal resetting rate $r^*$ that makes the search process most efficient~\cite{SMdiff,SMopt,SM2013,Whitehouse2013,Montero2013,EM14,Kusmierz2014,RUK14,KG15,RRU15,CS15,PKE16,Reuveni16,PR17,MMV17,BEM17,CS18,Belan18,PP19}. There is yet another interesting aspect of resetting dynamics in addition to 
optimising the search process. The resetting move breaks detailed balance and  
hence drives the system into a nontrivial non-equilibrium stationary state (NESS). Characterising such a NESS
has recently become a problem of central interest in statistical physics~\cite{SMdiff,Pal15,MSS15a,FGS16,NG16,MT17,FCBGM17}.

\begin{figure}
\includegraphics[width=\linewidth]{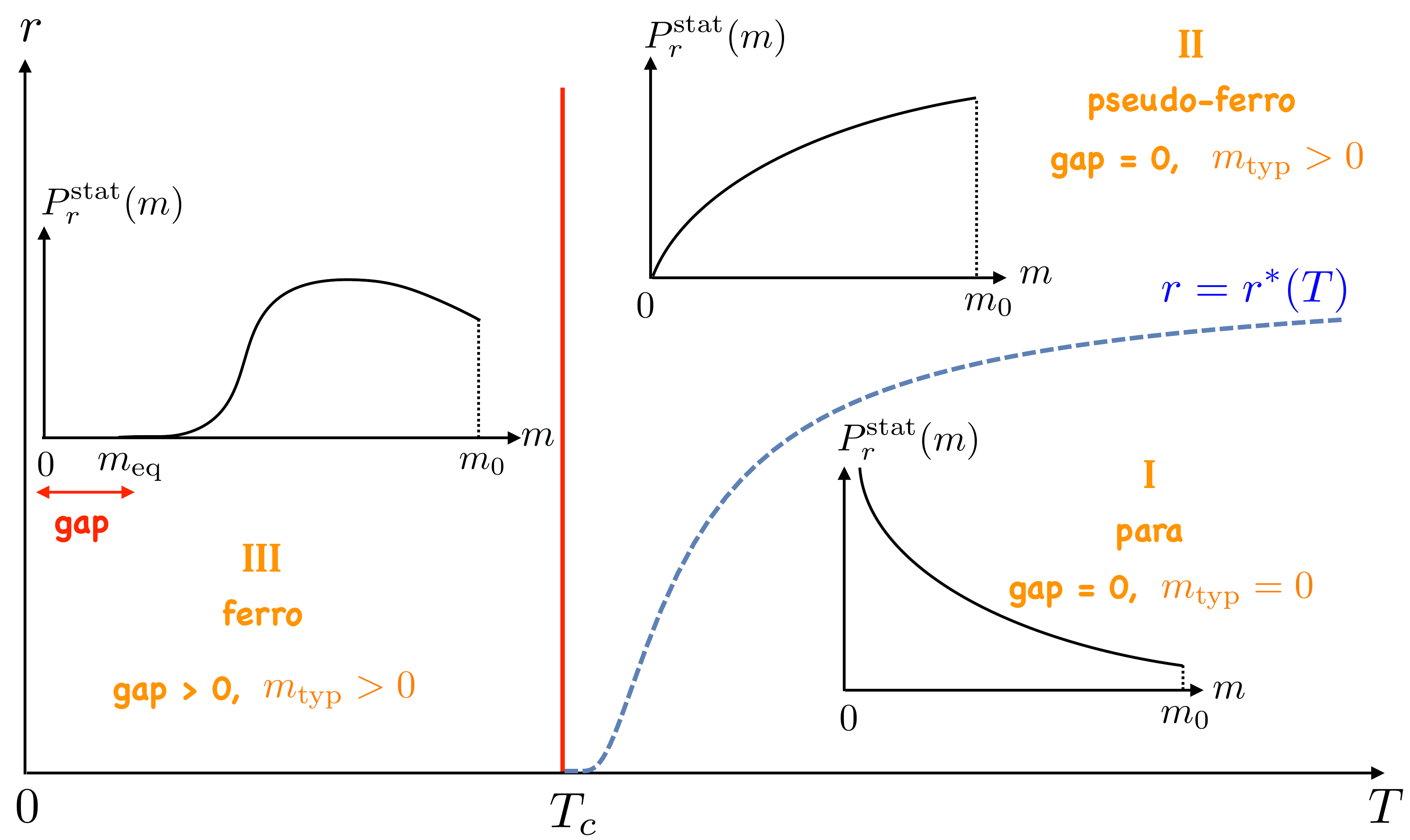}
\caption{(Color online) Phase diagram in the $(T,r)$ plane. The magnetisation distribution $P_r^{\rm stat}(m)$ in the stationary state, in the presence of resetting, is shown schematically in three different regions in the $(T,r)$ plane. For $T>T_c$, there is a crossover line $r^*(T)$ (shown schematically by the dashed blue line) that separates the para phase for $r<r^*(T)$ and the ``pseudo-ferro'' phase for $r>r^*(T)$. In the para phase, the $P_r^{\rm stat}(m)$ has a divergent peak at $m=m_{\rm typ}=0$ (and no gap at $m=0$), while in the pseudo-ferro phase, $P_r^{\rm stat}(m)$ vanishes at $m=0$ with a peak at $m_{\rm typ} >0$ (and still no gap at $m=0$). For $T<T_c$ (ferro phase), a nonzero gap opens up at $m=0$ in $P_r^{\rm stat}(m)$ and moreover the distribution peaks at $m_{\rm typ} > 0$.}\label{Fig_PhDiag}
\end{figure}
The phenomenon of resetting has found a large number of applications across disciplines. For example, in biology, the process of RNA polymerisation, which is responsible for the synthesis of RNA from a DNA template, is stochastically interrupted by backtracking \cite{Roldan2016,Lisica2016}. Similar notions are also found in the ecological context: for instance, animals such as rhesus monkeys \cite{GM05,GM06}, during the foraging period, are known to perform stochastic resetting to the previously visited sites and the effects of
such memory induced resetting have been studied in several models~\cite{BSS14,BEM17,FCBGM17,MSS15b,BR14, BP16}. In computer science, stochastic restarts can be used to reduce the running time of randomised search algorithms \cite{VAVA91,LSZ93,Montanari2002,TFP08,APZ13,Lorenz18}. Stochastic resetting has also been studied in the context of complex chemical processes as in the Michaelis-Menton reaction scheme \cite{RUK14, RRU15}, active run-and-tumble particles \cite{Scacchi2017,EM18,Mas19}, biological traffic models \cite{GMKC18}, and also recently in quantum systems~\cite{Mukherjee2018,Rose2018}.

Most of the systems discussed above concern the effect of resetting on the statics and dynamics of
a single particle (or equivalently for noninteracting systems). It is natural to ask how resetting affects 
the stationary state of a many-body interacting system. This question has been addressed in a number 
of interacting systems in one dimension, such as reaction-diffusion systems \cite{Durang2014}, fluctuating interfaces \cite{Gupta2014}, exclusion and zero-range processes subjected to resetting \cite{BKP19,KN20,Gr19}. However, none of these systems, in equilibrium, exhibit a thermodynamic phase transition. 
It is then interesting to know how resetting affects a system that, in the absence of resetting, displays a phase transition in its
equilibrium state. The simplest paradigmatic model that exhibits an equilibrium phase transition is the Ising model in $d$-dimensions, with
$d \geq 2$. It is then natural to ask the question: what kind of stationary state is reached if an Ising model, evolving under the natural
Glauber dynamics, is subject to resetting (to its initial configuration) at a constant rate~$r$?

It is useful to first recall the properties of the nearest neighbour Glauber Ising model in the absence of resetting ($r=0$) \cite{Glauber1963}. Let $s_i = \pm 1$ denote the spin at site $i$ of the Ising model. Starting from an initial configuration where the spins are independently chosen to have value $\pm 1$
with probability $(1\pm m_0)/2$ (where $m_0 \in [0,1]$), the individual spins flip according to the
Glauber rate that satisfies detailed balance (see later for details). Let $P(\{ s_i\},t)$ denote the probability distribution of a spin
configuration $\{s_i\}$ at time $t$. The most natural observable is the order parameter, i.e., the average 
magnetisation $m(t) = \frac{1}{N}\sum_i \langle s_i (t)\rangle$ where $\langle \cdots \rangle$ denotes averaging over the
probability measure $P(\{ s_i\},t)$ at time $t$. At late times, the system approaches the thermal equilibrium state, where the magnetisation
$m(t)$ approaches, irrespective of the initial value $m_0$, the final value $m_{\rm eq}>0$ for $T<T_c$ (ferromagnetic phase), and $0$
for $T \geq T_c$ (paramagnetic phase and at the critical point). Of course, $T_c=0$ in $d=1$.

What happens to the magnetisation $m(t)$ at long times when a finite resetting rate $r$ is switched on? This means that the system still evolves
under the Glauber dynamics but, with a rate $r$, it now goes back to the initial configuration and the Glauber dynamics restarts. In this Letter
we show that this nonzero resetting drives the system into a new non-equilibrium stationary state where the magnetisation has a nontrivial distribution and not a single value as in thermal equilibrium. This nontrivial stationary state arises from the fact that, even though after each resetting event the system starts the Glauber dynamics from the same initial configuration, the state of the system at time $t$ is governed by the time of evolution since the last resetting event and this time is itself a random variable drawn from an exponential distribution with mean $1/r$. 
Consequently, the measured thermally averaged magnetisation $m(t)$ fluctuates from one resetting history to another. In particular, in the stationary state it acquires a nontrivial distribution. This is thus markedly different from the equilibrium case ($r=0$) where the magnetisation distribution is trivially a delta function centred either at $m=0$ (for $T>T_c$) or at $m=m_{eq}>0$ (for $T<T_c$). Thus the knowledge of the full distribution of the magnetisation is necessary to characterise the steady state of the system in the presence of resetting.

It is useful to first summarise our main results. We show that a nonzero resetting leads to a rich phase diagram in the $(T,r)$ plane as displayed in Fig. \ref{Fig_PhDiag} with the emergence of a new phase which we call ``pseudo-ferro'' phase. At all temperatures, the stationary magnetisation distribution has now a finite support. 
For $T>T_c$, there is a new crossover line $r^*(T)$ that separates the paramagnetic phase $r<r^*(T)$ from a pseudo-ferro phase
for $r>r^*(T)$. In the para phase, the stationary magnetisation distribution $P^{\rm stat}_r(m)$ {\it diverges} as $m \to 0$, as $P^{\rm stat}_r(m) \sim m^{\zeta}$ with
$\zeta = r/r^*(T) - 1 < 0$ and hence the typical magnetisation $m_{\rm typ} =0$ ($m_{\rm typ}$ denotes the value of $m$ at which $P^{\rm stat}_r(m)$ reaches its maximum). In addition, there is no gap at $m=0$, i.e. $g=0$. In contrast, in the ``pseudo-ferro'' phase ($T>T_c$ and $r > r^*(T)$), while the gap $g$ still remains zero, $P^{\rm stat}_r(m)$ now vanishes as $m \to 0$ as $P^{\rm stat}_r(m) \sim m^{\zeta}$ with $\zeta = r/r^*(T) - 1 > 0$. Consequently, the maximum of $P^{\rm stat}_r(m)$ occurs at a nonzero value $m_{\rm typ} > 0$ (see Fig. \ref{Fig_PhDiag}). For $T<T_c$ (ferro phase), the distribution $P^{\rm stat}_r(m)$ has a finite support $[m_{\rm eq},m_0]$ (for $m_{\rm eq} < m_0$) or over $[m_0, m_{\rm eq}]$ (if $m_{\rm eq} > m_0$). Thus, in this phase, there
is a finite nonzero gap $g= \min(m_{\rm eq},m_0)$. In addition, $m_{\rm typ} >0$ in the ferro phase. Exactly at $T=T_c$, the distribution $P_r^{\rm stat}(m)$ vanishes extremely rapidly, $P^{\rm stat}_r(m) \sim e^{- A m^{-\kappa}}$ as $m \to 0$. We show that the exponent $\kappa$ is related to the equilibrium critical exponents via the relation $\kappa = \nu z/\beta$, where $\nu$ and $\beta$ are respectively the correlation length and the order parameter critical exponents, while $z$ is the dynamical critical exponent associated to the Ising Glauber dynamics at $T=T_c$. We establish these results from an exact solution in $d=1$ and, for $d=2$, we provide scaling arguments supported by numerical simulations.

\begin{figure*}[t]
\includegraphics[width = \linewidth]{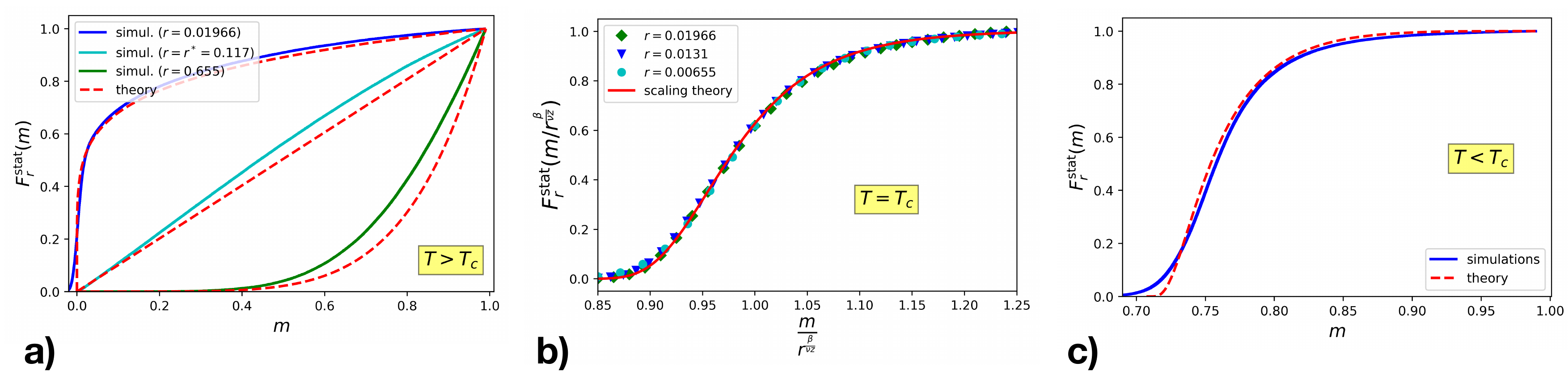}
\caption{(Color online) Plot of the CDF $F_r^{\rm stat}(m) = \int_0^m P_r^{\rm stat}(m')\,dm'$ vs $m$, for different values of the resetting rate $r$, in {\bf a)} the para phase $T>T_c$, {\bf b)} at the critical point $T=T_c \approx 2.269$ (for $J=k_B=1$) and {\bf c)} in the ferro phase $T<T_c$. {\bf a)}:  the solid curves show the simulation results for $T=3.5$, $m_0 = 0.9905$ and for three different  values of $r$, respectively for $r<r^*$ (top blue curve), $r = r^* \approx 0.117$ (middle magenta curve) and $r>r^*$ (bottom green curve). The red dashed lines correspond to our theoretical prediction obtained by integrating Eq.~(\ref{eq:parstat}). The agreement with the theoretical predictions gets better for smaller $r$ as explained in the text. {\bf b)}: scaling collapse at $T=T_c$ of the CDF $F_r^{\rm stat}(m) \approx H(m \,r^{-\beta/(\nu z)})$, compared with the theoretical function $H(y) = e^{-(b_c/y)^{\nu z/\beta}}$ (red solid line) for three different small values of $r$. Here we used $\beta=1/8$, $\nu = 1$ and $z \approx 2.17$ together with the estimated parameter $b_c = 0.9576$. {\bf c)}: CDF for $T=2.24<T_c$ with $m_{eq} = 0.70732$ and $r=0.00459$ (blue solid line), compared with the theoretical prediction (see the text and Eq. (23) in the Supp. Mat. \cite{supp_mat}) shown by the dashed red line. All the simulations were performed on a $256 \times 256$ square lattice. One observes finite size effects at the edges of the support.} \label{Fig_numerics}
\end{figure*}

We start with an Ising model with ferromagnetic nearest-neighbour interactions $H = - J \sum_{\langle i,j \rangle} s_i s_j$ on a $d$-dimensional lattice with $N$ sites and periodic boundary conditions. Starting from an initial condition where the spins are independently $\pm 1$ with probability $(1 \pm m_0)/2$, the Glauber dynamics, in the absence of resetting, consists in flipping a single spin with rate \cite{Glauber1963}
\begin{eqnarray}\label{rate}
w(s_i \to - s_i) = \frac{1}{1+e^{\beta \Delta E}} \;,
\end{eqnarray} 
where $\beta = 1/(k_BT)$ is the inverse temperature and $\Delta E = 2 J s_i \sum_{j \in {\rm n. n.}} s_j$ is the change of energy in flipping the 
$i$-th spin. In $d=1$, this rate simplifies to $w(s_i \to - s_i)  = (1/2) (1 - \gamma s_i(s_{i-1} + s_{i+1})/2)$ where $\gamma = \tanh{(2 \beta J)}$. This property makes the $1d$ Glauber dynamics exactly solvable as the evolution equation for the $n$-point correlation functions only involve $n$-point functions, i.e. it satisfies a closure property \cite{Glauber1963}. This closure property however does not hold for $d>1$. Note that under the Glauber dynamics, the magnetisation $m(t) = (1/N) \sum_i \langle s_i(t) \rangle$ evolves deterministically with time $t$. 

Now imagine that we 
switch on the resetting mechanism, whereby the system goes back randomly in time to the initial configuration with a nonzero rate $r$.    
This means that, between two successive resetting events, the system evolves by the standard Glauber dynamics mentioned above (\ref{rate}). 
If we now observe the system at a fixed time $t$, what matters is the time $\tau$ elapsed since the last resetting before $t$. This is
because the system has evolved by the pure Glauber dynamics during the interval $[t-\tau,t]$. But since the resettings happen stochastically, the time $\tau$ itself is a random variable. As a result, any observable, such as the average magnetisation, measured at time $t$ also becomes a random variable. One can express the distribution $P_r(m,t)$ of the average magnetisation $m$ in the presence of resetting with rate $r$ by the simple renewal equation
\begin{equation}\label{renewal1}
P_r(m,t) = r \int_0^t d\tau \, e^{- r \tau} \, P_0(m,\tau) + e^{-r t} \, P_0(m,t) \;,
\end{equation}  
where $P_0(m,\tau) = \delta(m-m(\tau))$ denotes the magnetisation distribution in the absence of resetting ($r=0$) since it evolves deterministically  as $m(\tau)$. The second term in~(\ref{renewal1}) is easy to explain: it corresponds to having no resetting up to time $t$ and the system evolves by the standard Glauber dynamics during $[0,t]$ and ends up with a magnetisation $m$ at time $t$. The first term in (\ref{renewal1}) corresponds to the event that there is a resetting event at time $t-\tau$ which happens with probability $r\, d\tau$, followed by no-resetting in the interval $[t-\tau,t]$ which occurs with probability $e^{-r \tau}$. During this interval of length $\tau$, the system evolves via the standard Glauber dynamics, hence at time $t$, the magnetisation is just $m(\tau)$. In the large time limit, the second term in (\ref{renewal1}) drops out and hence the stationary magnetisation distribution is given by 
\begin{equation}
P_r^{\text{stat}}(m) = r \int_0^{\infty} d\tau \, e^{-r\tau} \, \delta(m-m(\tau)) \, \;.
\label{trick}
\end{equation} 
Given this simple renewal property (\ref{trick}), we need just to know the deterministic Glauber evolution $m(\tau)$ for all $\tau$ in the absence of resetting to determine $P_r^{\text{stat}}(m)$.

\textit{Resetting in the $1d$ Ising model.} We start with the exactly solvable case on a $1d$-lattice with $N$ sites and periodic boundary conditions in the absence of resetting. The one-point average $\langle s_i(t) \rangle$ evolves via \cite{Glauber1963}
\begin{eqnarray}\label{Glauber_mag}
\frac{d}{dt} \langle s_i(t) \rangle = - \langle s_i(t) \rangle + \frac{\gamma}{2}(\langle s_{i-1}(t) \rangle + \langle s_{i+1}(t) \rangle) \;,
\end{eqnarray}
with initial condition $\langle s_i(0) \rangle = m_0$ for all $i$. Thus the average magnetisation $m(t) = (1/N) \sum_i \langle s_i(t) \rangle$ evolves via $dm(t)/dt = -(1-\gamma ) m(t)$ whose solution is trivially 
 \begin{equation}
m(t) = m_0 \, e^{-(1-\gamma) t} \;; \;\quad \gamma = \tanh{(2 \beta J)} \;.
\label{eq:decay}
\end{equation}
Substituting this solution (\ref{eq:decay}) in Eq. (\ref{trick}), one obtains exactly
\begin{eqnarray}\label{PDF_1d}
P^{\rm stat}_r(m) = \frac{r}{m_0(1-\gamma)} \left( \frac{m}{m_0}\right)^{\frac{r}{r^*(T)}-1}  \;, \; m \in [0,m_0] \;
\end{eqnarray}
where 
\begin{eqnarray}\label{rstar_1d}
r^*(T) = 1 - \gamma = 1 - \tanh\left(\frac{2J}{k_B T}\right) \;.
\end{eqnarray}
In this case $T_c=0$ and we have only the part of the phase diagram in Fig. \ref{Fig_PhDiag} with $T \geq T_c$. The result in Eq.~(\ref{PDF_1d}) clearly shows that, near $m=0$, $P^{\rm stat}_r(m) \sim m^{\zeta}$ where the exponent 
$\zeta = r/r^*(T)-1$ varies continuously with temperature. Thus $P^{\rm stat}_r(m)$ either diverges (for $r<r^*(T)$) or vanishes (for $r>r^*(T)$) as $m \to 0$. In the former case, $m_{\rm typ} =0$ -- this is the para phase. In contrast, for $r>r^*(T)$, $m_{\rm typ} >0$: this is the new ``pseudo-ferro'' phase induced by resetting. We have also done numerical simulations in $d=1$ to verify 
our analytical prediction in Eq. (\ref{PDF_1d}) and found excellent agreement (see Fig. 1 in the Supplementary Material~\cite{supp_mat}).

\noindent
\textit{Resetting in the $2d$ Ising model.} In $2d$, the Ising model at equilibrium has a finite $T_c \approx 2.269$ with the choice $J=k_B=1$. 
Unlike in $d=1$, the Glauber dynamics is not exactly solvable in $d=2$. However, using the well established phenomenological behaviour of $m(t)$, in particular at late times, in the renewal equation (\ref{trick}), we can make some predictions for the stationary magnetisation distribution $P^{\rm stat}_r(m)$ in various parts of the phase diagram in the $(T,r)$ plane in Fig. \ref{Fig_PhDiag}. We then verify these predictions with numerical simulations and find a very good agreement. Below we consider the three cases $T>T_c$, $T<T_c$ and $T=T_c$ separately.


{\it $T>T_c$}. We start with the paramagnetic phase $T>T_c$. In this case, the average magnetisation, for the pure Glauber dynamics is expected to decay at late times as $m(t) \sim a_1\, e^{-\lambda_1 \, t}$ where the amplitude $a_1$ and the leading decay rate $\lambda_1$ both depend on temperature \cite{Stauffer1997} and can be estimated very precisely from Monte Carlo simulations. This pure exponential decay of $m(t)$ holds only when $t \gg 1/\Delta \lambda$ where
$\Delta \lambda$ is the first gap in the relaxation spectrum. Substituting this functional form in Eq. (\ref{trick}) we get, for $r \ll \Delta \lambda$
\begin{equation}
P_r^{\text{stat}}(m) \approx \frac{r}{\lambda_1 a_1^{\frac{r}{\lambda_1}}} m^{\frac{r}{\lambda_1} -1}, \quad \quad m \in (0,a_1] \;.
\label{eq:parstat}
\end{equation}
This is a good approximation at high temperature $T \gg T_c$ where $\Delta \lambda$ is large. In this case $a_1 \approx m_0$ and thus we recover qualitatively a similar distribution for $P^{\rm stat}_r(m)$ (\ref{PDF_1d}) as in the $d=1$ case. In the case of $2d$, for $T \gg T_c$, we then have $r^*(T) = \lambda_1$. Thus, as in the $d=1$ case, we have a crossover from the usual para phase for $r < r^*(T)$ to the ``pseudo-ferro'' phase for $r>r^*(T)$ across the crossover line $r^*(T)$ in the $(T,r)$ plane, for $T \gg T_c$. Our numerical simulations are completely consistent with this scenario. In Fig. \ref{Fig_numerics} a), we plot the cumulative stationary distribution $F^{\rm stat}_r (m) = \int_{0}^m P^{\rm stat}_r(m') dm'$ vs $m$ for three different resetting rates: one in the para phase (the top curve), one in the ``pseudo-ferro'' phase (bottom curve) and finally at the crossover line $r = r^*(T)$ (middle curve). In the last case, the probability distribution function (PDF) of the magnetisation $P^{\rm stat}_{r^*}(m)$ is uniform [from Eq. (\ref{eq:parstat})] and hence the cumulative distribution function (CDF) $F^{\rm stat}_{r^*} (m)$ increases linearly with $m$.

\textit{$T < T_c$.} In the ferro phase and in the absence of resetting, the magnetisation density of the $2d$ Ising model reaches a nonzero equilibrium value $m_{eq}$ with a stretched exponential decay at late times \cite{Stauffer1997}
\begin{equation}
m(t) \approx m_{eq} \pm a e^{-bt^c} \;,
\label{evferr}
\end{equation}
where the parameters $a, b$ and $0<c<1$, not known analytically, need to be determined from simulations. In (\ref{evferr}), the $+$ and $-$ signs are used in the case $m_0 > m_{eq}$ and $m_0 < m_{eq}$ respectively. When a constant resetting rate $r$ is introduced in the system, the stationary PDF $P_r^{\text{stat}}(m)$ is obtained from the general formula in Eq.~(\ref{trick}), where for $m(\tau)$ we now use Eq. (\ref{evferr}). The resulting $P_r^{\rm stat}(m)$ is non-trivial and its detailed form is discussed in \cite{supp_mat}. For instance, a plot of the associated CDF is given in Fig. \ref{Fig_numerics} c) for the case $m_0 > m_{eq}$ and compared to simulations, showing an excellent agreement. In this case, the PDF is supported over the finite interval $[m_{eq}, m_{eq} + a]$. It has non-trivial asymptotic behaviours at the edges. For example, near the lower edge, where $m \to m_{eq}^+$, $P_r^{\rm stat}(m)$ vanishes faster than a power law as $P_r^{\rm stat}(m) \sim \exp(- B [-\ln(m-m_{eq})]^{1/c})$ where $0<c<1$ and $B =r\,b^{-1/c}$. 

\textit{$T=T_c$.} Exactly at the critical point, the magnetisation $m(t)$, without resetting, has a non-monotonic decay with time \cite{Janssen,Zheng1998,CG05}, which is not known analytically, except at short and long times. Hence it is difficult to evaluate $P_r^{\rm stat}(m)$ exactly from Eq. (\ref{trick}) for all $m$. However, when $r$ and $m$ are both small, one can use in Eq. (\ref{trick}) the late time form of $m(t) \approx b_c \, t^{-\phi}$ where the exponent $\phi=\beta/(\nu z)$ is related to the standard critical exponents defined earlier. We then find that there is a scaling regime as $m \to 0$, $r \to 0$ but with $m \, r^{-\phi}$ fixed where the distribution $P_r^{\rm stat}(m)$ takes the scaling form 
\begin{eqnarray}\label{scaling_Tc}
P_r^{\rm stat}(m) \approx r^{-\phi} G(m \, r^{-\phi}) \;,
\end{eqnarray}
where the scaling function~$G(y) = A\, y^{-1 - 1/\phi} e^{-(b_c/y)^{1/\phi}}$, with $A = b_c^{1/\phi}/\phi$, vanishes extremely rapidly as $y \to 0$. Consequently the CDF $F_r^{\rm stat}(m) = \int_0^m P_r^{\rm stat}(m')\,dm' \approx H(m \, r^{-\phi})$ where $H'(y) = G(y) = e^{-(b_c/y)^{1/\phi}}$. This scaling behaviour is verified numerically in Fig. \ref{Fig_numerics} b) where $F_r^{\rm stat}(m)$ shows a beautiful scaling collapse for three different values of $r$. The full distribution $P_r^{\rm stat}(m)$ has still a finite support $m \in [0, \Gamma]$ where the upper cut-off $\Gamma$ depends on system parameters. While there is no strict gap at $m=0$ (as in the ferro phase), $P_r^{\rm stat}(m)$ vanishes extremely rapidly as $m \to 0$. Thus, even in this resetting induced NESS, there is a remnant signature of the equilibrium critical point $T_c$ that is manifest in this essential singularity near $m=0$.

To summarise, in this Letter, we have addressed a general question: how does resetting affect a many-body interacting system that, in equilibrium, exhibits a thermodynamic phase transition at $T=T_c$? A natural candidate to study this question is the paradigmatic Ising model evolving under the Glauber dynamics. We have shown, both analytically and numerically, that resetting (with a constant rate $r$) leads to a nontrivial phase digram of the Glauber-Ising model in the $(T,r)$ plane (see Fig. \ref{Fig_PhDiag}). In particular, we have shown that resetting leads to the emergence of a new pseudo-ferro phase for $T>T_c$ and $r > r^*(T)$ where the system has a non-zero typical magnetisation in the stationary state and yet there is no gap in the magnetisation distribution near $m=0$. The qualitative features of the phase diagram, established here for the $d=1$ and $d=2$ Ising model with Glauber dynamics, are also expected to hold in higher dimensions, as well as for other single spin-flip dynamics, such as the Metropolis dynamics. Finally, going beyond the magnetisation distribution, it would be interesting to investigate the structure of the distribution of the two-point correlation functions in this resetting induced non-equilibrium stationary state.



\end{document}


\title{Ising model with stochastic resetting: supplemental material}

\author{Matteo Magoni$^1$$^{,}$$^2$, Satya N. Majumdar$^1$, and Gr\'egory Schehr$^1$}

\affiliation{$^1$Universit\'e Paris-Sud, CNRS, LPTMS, UMR 8626, 91405 Orsay, France \\
$^2$Department of Applied Science and Technology, Politecnico di Torino, Corso Duca degli Abruzzi 24, 10129 Torino, Italy}

\date{\today}

\begin{abstract}
We give the principal details of the computations and simulations described in the main text of the Letter.
\end{abstract}


\maketitle 

\section{Resetting in the 1$d$ Ising model: Monte Carlo simulations}

For the $1d$ Ising model, $T_c = 0$ and hence, there is no ferromagnetic phase for $T > 0$. Hence, in the phase diagram of Fig. 1 of the main text, we have only the phase $T \geq T_c = 0$. The stationary probability distribution of the average magnetisation in the 1$d$ Ising model in the presence of resetting is given by Eq. (6) in the main text and reads
\begin{eqnarray}\label{PDF_1d}
P^{\rm stat}_r(m) = \frac{r}{m_0(1-\gamma)} \left( \frac{m}{m_0}\right)^{\frac{r}{r^*(T)}-1}, \qquad \qquad m \in [0,m_0] \;
\end{eqnarray}
where 
\begin{eqnarray}\label{rstar_1d}
r^*(T) = 1 - \gamma = 1 - \tanh\left(\frac{2J}{k_B T}\right) \;.
\end{eqnarray}

\begin{figure*}[h]
\centering
\includegraphics[width = \linewidth]{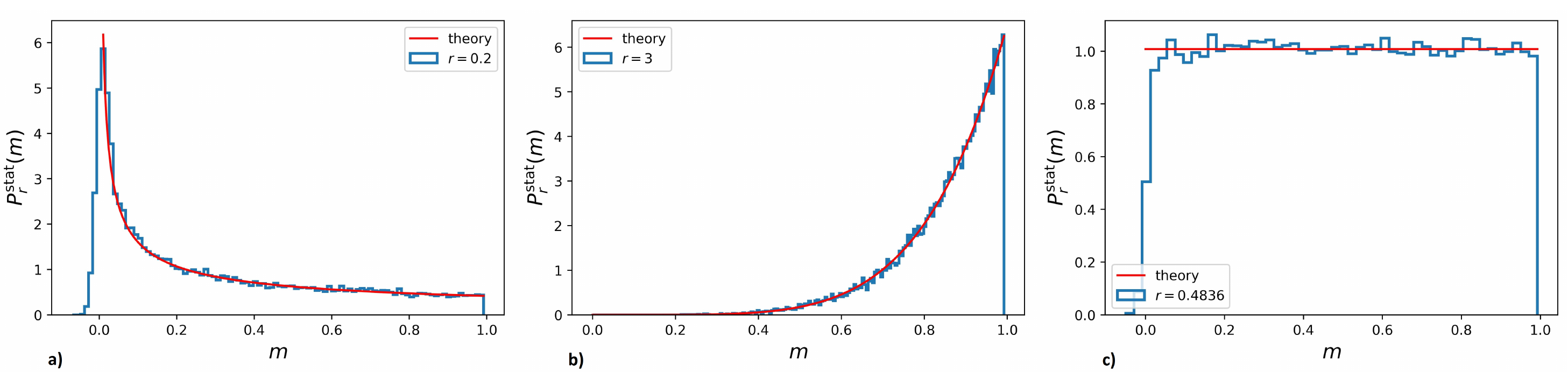}
\caption{Plot of the PDF $P^{\rm stat}_r(m)$ for three different values of $r$: (a) para phase with $r = 0.2 < r^*(T=3.5) = 0.4836\ldots$, (b) pseudo-ferro phase with $r=3>r^*(T=3.5)$ and (c) exactly on the crossover line $r=r^*(T=3.5)$. The red solid lines correspond to the theoretical result in Eq. (\ref{PDF_1d}), while the blue curves correspond to numerical simulations. The almost blue vertical lines at the edges of the distribution are due to finite size effects.}
\label{fig:totsimul}
\end{figure*}

We have performed Monte-Carlo simulations on a one-dimensional lattice of length $N=10000$ at temperature $T=3.5$ (setting $J=k_B=1$) to verify our analytical results. Simulation data are collected by averaging over 3000 independent realisations. The fixed initial configuration to which the system is reset at a constant rate $r$ has magnetisation $m_0 = 0.992$. Fig.~\ref{fig:totsimul} presents the plot of the PDF $P^{\rm stat}_r(m)$ for three different values of $r$, showing the excellent agreement between theory (red line) and Monte-Carlo simulations (plotted in blue).

We perform the Glauber dynamics of the $1d$ Ising model at temperature $T=3.5$, for which $r^*(T=3.5)=1 - \tanh(2/3.5) = 0.4836$. Hence, for $r < r^*(T=3.5)$, $P^{\rm stat}_r(m)$ diverges as $m \rightarrow 0$. This is clearly shown in Fig. 1 a), where, in the presence of a resetting rate $r=0.2$, the typical value of the average magnetisation is $m_{\rm typ} =0$: this regime is the para phase. In contrast, for $r > r^*(T=3.5)$, a new ``pseudo-ferro'' phase emerges, where $m_{\rm typ} > 0$ and $P^{\rm stat}_r(0) = 0$, as shown in Fig. 1 b). Exactly on the crossover line $r=r^*(T)$, Eq. (\ref{PDF_1d}) predicts a uniform distribution for $P^{\rm stat}_r(m)$ in $[0,m_0]$: this is numerically verified in Fig. 1 c).

\begin{figure}[h]
  \centering
  \includegraphics[width=.45\linewidth]{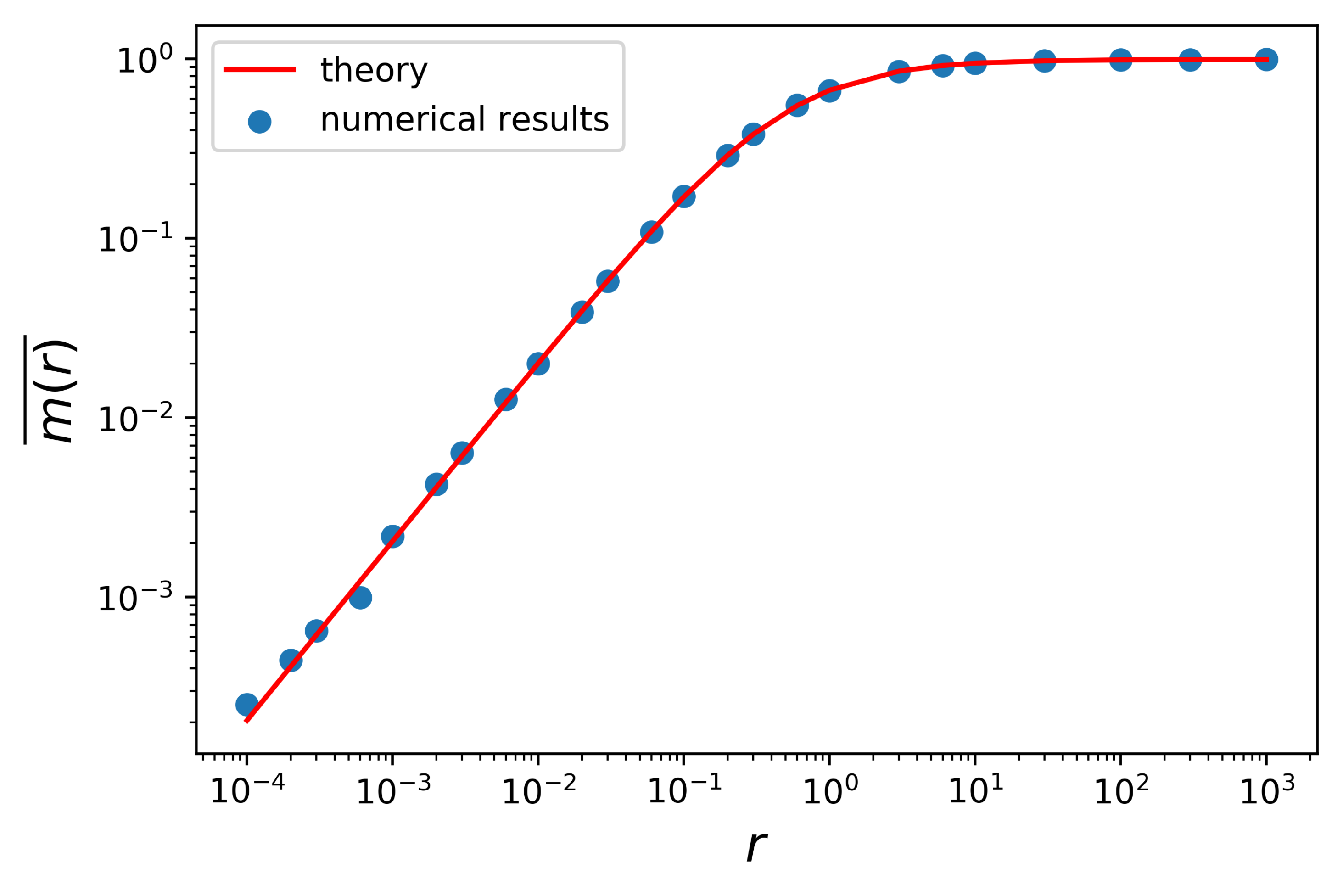}  
  \caption{Plot of the first moment $\overline{m(r)}$ as a function of $r$. Numerical data and Eq.~(\ref{eq:avg_1D}) are plotted in blue and red respectively.}
  \label{fig:1Davg}
\end{figure}

It is also interesting to know how the first moment of $P^{\rm stat}_r(m)$ depends on the resetting rate $r$. This can be computed easily as
\begin{equation}
\overline{m(r)} = r \int_0^{\infty} d\tau \, e^{-r\tau} \, m_0 \, e^{-(1-\gamma) t} = \frac{r}{r+r^*(T)} \, m_0 \;,
\label{eq:avg_1D}
\end{equation}
where $r^*(T) = 1 - \gamma$.
We have measured this quantity $\overline{m(r)}$ for different values of $r$ from our numerical simulations. They are plotted in blue in Fig.~\ref{fig:1Davg}, showing an excellent agreement with Eq.~(\ref{eq:avg_1D}). However, as discussed in the main text, the average magnetisation is not enough to characterise the stationary state under resetting and one needs the full distribution $P^{\rm stat}_r(m)$.

\section{Scaling predictions for $P_r^{\text{stat}}(m)$ in the three temperature regimes for the Ising model with resetting in $d \geq 2$}


For the Ising model in $d\geq 2$, the critical temperature $T_c$ is finite. Hence, in the phase diagram of Fig. 1 of the main text, we will have all the three
phases. Unlike in $1d$, the Glauber dynamics is not exactly solvable in $d \geq 2$, even in the absence of resetting.  
In this section, we provide some general scaling arguments to predict scaling forms for   
$P_r^{\text{stat}}(m)$ for $T>T_c$, $T<T_c$ and $T=T_c$, using the renewal equation (3) in the main text which reads
\begin{equation}
P_r^{\text{stat}}(m) = r \int_0^{\infty} d\tau \, e^{-r\tau} \, \delta(m-m(\tau)) \, \;.
\label{trick_sup}
\end{equation} 

\subsection{The high temperature phase $T>T_c$}
In the high temperature phase, the average magnetisation, quite generally, decays with time as $m(t) = \sum_{i} a_i e^{-\lambda_i t}$ where $\lambda_1< \lambda_2 < \cdots$ denote the eigenvalues associated to the relaxation spectrum. At late times, this is dominated by the lowest eigenvalue $\lambda_1$, i.e.   
$m(t) \approx a_1\, e^{-\lambda_1 \, t}$. This pure exponential decay is thus valid when $t \gg 1/\Delta \lambda$ where $\Delta \lambda = \lambda_2 - \lambda_1$ is the first gap in the relaxation spectrum. Substituting this pure exponential form in Eq. (\ref{trick_sup}) we get, for $r \ll \Delta \lambda$
\begin{equation}
\begin{split}
P_r^{\text{stat}}(m) &\approx r \int_0^{\infty} d\tau \, e^{-r\tau} \, \delta(m-a_1\, e^{-\lambda_1 \, \tau}) = \\
&\approx \frac{r}{\lambda_1 a_1^{\frac{r}{\lambda_1}}} \, m^{\frac{r}{\lambda_1} -1}, \quad \quad m \in (0,a_1] \;.
\end{split}
\end{equation}
This is Eq. (8) of the main text and is valid for $T \gg T_c$, such that the first gap $\Delta \lambda$ is large.

\subsection{The low temperature phase $T<T_c$}
In the low temperature phase, the average magnetisation $m$ approaches the equilibrium magnetisation $m_{eq} > 0$ as a stretched exponential at late times \cite{Stauffer1997}, i.e., $m(t) \approx m_{eq} \pm a \, e^{-b t^c}$, where the $+$ and $-$ signs are used in the cases $m_0 > m_{eq}$ and $m_0 < m_{eq}$ respectively.

Inserting this expression for $m(t)$ in Eq. (\ref{trick_sup}) of the main text, we obtain for $r \ll \Delta \lambda$
\begin{equation}
\begin{split}
P_r^{\text{stat}}(m) &\approx r \int_0^{\infty} d\tau \, e^{-r\tau} \, \delta(m - m_{eq} \mp a \, e^{-b \tau^c})  \\
&\approx \frac{r}{b \, c} \int_0^a dx \, \frac{1}{\left[\frac{1}{b} \ln\left(\frac{a}{x}\right)\right]^{\frac{c-1}{c}} \, x} \, \exp\left\{-r\left[\frac{1}{b} \ln\left(\frac{a}{x}\right)\right]^{\frac{1}{c}}\right\} \, \delta(m-m_{eq} \mp x)  \\
&\approx \frac{r}{b \, c \, [f(m)]^{\frac{c-1}{c}} \, |m - m_{eq}|} \, \exp\left\{-r \, [f(m)]^{\frac{1}{c}}\right\} \; , 
\end{split}
\end{equation}
where $f(m) = \frac{1}{b} \ln\big(|\frac{a}{m - m_{eq}}|\big)$. Its support is $(m_{eq},m_{eq} + a]$ if $m_0 > m_{eq}$ or $[m_{eq}-a,m_{eq})$ if $m_0 < m_{eq}$. In the second equality we have used $x = a \, e^{-b \tau^c}$.

\subsection{At the critical point $T=T_c$}
At $T=T_c$, the late time evolution of $m(t)$ in the absence of resetting follows a power law decay to 0 since the spectrum is gapless. In this case, at late times, $m(t) \approx b_c \, t^{-\phi}$ where the exponent is $\phi=\beta/(\nu z)$ \cite{CG05}. More precisely, this power law decay holds for $t < L^z$, where $L$ is the linear size of the system. This condition is fulfilled in our simulations since $\frac{1}{r_{\text{min}}} \ll L^z$, where $r_{\text{min}}$ is the smallest resetting rate used in our simulations. Using Eq. (\ref{trick_sup}) of the main text, we obtain for $r \ll 1$
\begin{equation}
\begin{split}
P_r^{\text{stat}}(m) & \approx r \int_0^{\infty} d\tau \, e^{-r\tau} \, \delta(m - b_c \, t^{-\phi}) \\
&\approx r \, \frac{ b_c^{\frac{1}{\phi}}}{\phi} \int_0^{\infty} dx \, \frac{1}{x^{\frac{\phi+1}{\phi}}} \exp\left[-r\left(\frac{b_c}{x}\right)^{\frac{1}{\phi}}\right] \, \delta(m-x) \\
& \approx \frac{r \, b_c^{\frac{1}{\phi}}}{\phi \, m^{\frac{\phi+1}{\phi}}} \exp\left[-r\left(\frac{b_c}{m}\right)^{\frac{1}{\phi}}\right] \; ,
\end{split}
\label{eq:sol_crit}
\end{equation}
where in the second equality we made the change of variable $x = b_c \, t^{-\phi}$. In Eq. (\ref{eq:sol_crit}), if we set the scaling variable $y = m r^{-\phi}$, then $P_r^{\rm stat}(m)$ takes the scaling form given in Eq. (10) of the main text.

\section{Numerical estimate of the parameters in the time evolution $m(t)$ of the 2$d$ Ising model}

\begin{figure*}[h]
\centering
\includegraphics[width = \linewidth]{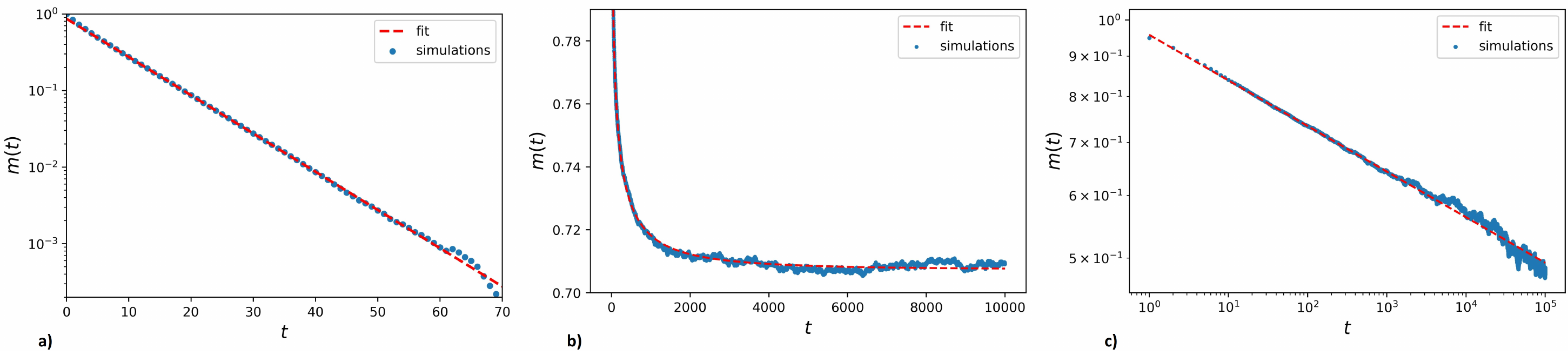}
\caption{Fit of $m(t)$ in the three temperature regimes.  For $T<T_c$ and $T=T_c$, it is necessary to consider much more time steps to obtain an accurate fit, since the decay to the equilibrium magnetisation is slower than the one for $T>T_c$.}
\label{fig:totevo}
\end{figure*}

In order to use Eq. (3) of the main text and compute $P_r^{\text{stat}}(m)$, we need to estimate the parameters in the time evolution $m(t)$. To this end, we have done Monte Carlo simulations of the Glauber dynamics in the three temperature regimes on a $256 \times 256$ square lattice with the choice $J = k_B = 1$ and with $m_0=0.9905$ as the magnetisation of the starting configuration at time $t=0$. Fig.~\ref{fig:totevo} shows the fit of $m(t)$ in the paramagnetic phase (Fig. 3 a)), ferromagnetic phase (Fig. 3 b)) and at the critical temperature (Fig. 3 c)). The blue dots (simulation data) are obtained by averaging over 4770, 490 and 70 independent realisations at temperatures $T=3.5$, $T=2.24$ and $T=T_c=2.269$ respectively. The red dashed lines are obtained by fitting the well established phenomenological behaviour of $m(t)$ in the three temperature regimes to the numerical data.  

As mentioned above, for $T > T_c$, the average magnetisation decays at late times as $m(t) \approx a_1 \, e^{-\lambda_1 t}$. The fit shown in Fig. 3 a) provides the values $a_1 = 0.889$ and $\lambda_1 = 0.117$ for $T=3.5$. For the low temperature phase, we set $T=2.24 < T_c = 2.269$ and in Fig. 3 b) we show the fit of the functional form $m(t) \approx m_{eq} + a \, e^{-b t^c}$ with estimated parameter values $a=0.283$, $b=0.37$ and $c=0.316$. At the critical temperature $T_c = 2.269$ the fit, $m(t) = b_c t^{-\phi}$, shown in Fig. 3 c) provides an estimation of the parameters $b_c = 0.9576$ and $\phi = 0.0576$. Using the relation $\phi = \beta/(\nu z)$, and the known exact values of $\beta = \frac{1}{8}$ and $\nu = 1$, one gets $z = 1/(8 \phi) \approx 2.17$, which is in excellent agreement with previous large-scale simulations \cite{Stauffer1997}.

\section{First moment of $P_r^{\rm stat}$ in the 2$d$ Ising model}

A quantity of interest that clearly depends on the resetting rate $r$ is the first moment of $P^{\rm stat}_r(m)$ in the 2$d$ Ising model. Similarly to the approach followed for the 1$d$ case, we can easily compute it for $T > T_c$, $T<T_c$ and $T=T_c$ as
\begin{equation}
\overline{m(r)} = r \int_0^{\infty} d\tau \, e^{-r\tau} \, m(\tau) \, , 
\label{eq:2D_avg_mag}
\end{equation}
where $m(t)$ is the time evolution of the average magnetisation in the Glauber dynamics which has been estimated in the previous section.

For $T > T_c$, using $m(t) \approx a_1 e^{-\lambda_1 t}$ at late times, the average magnetisation in the stationary state is given, for small $r$ (so that we can use the pure exponential late time decay of $m(t)$), by
\begin{equation}
\overline{m(r)} \approx r \int_0^{\infty} d\tau \, e^{-r\tau} \, a_1 \, e^{-\lambda_1 \tau} = \frac{r}{r + \lambda_1} \, a_1 \, ,
\label{eq:2D_avg_par}
\end{equation}
similarly to the one dimensional case Eq.~(\ref{eq:avg_1D}).

For $T < T_c$, using $m(t) \approx m_{eq} \pm a \, e^{-b t^c}$, $\overline{m(r)}$ takes the form, for small $r$, 
\begin{equation}
\overline{m(r)} = r \int_0^{\infty} d\tau \, e^{-r\tau} \, (m_{eq} + a \, e^{-b \tau^c}) = m_{eq} + r \, a \int_0^{\infty} d\tau \, e^{-r \tau - b \tau^c} \, ,
\label{eq:2D_avg_fer}
\end{equation}
where the last integral can be estimated numerically very accurately.

At the critical temperature $T_c$, using $m(t) \approx b_c t^{-\phi}$, the average magnetisation in the stationary state reads, for small $r$, 
\begin{equation}
\overline{m(r)} = r \int_0^{\infty} d\tau \, e^{-r\tau} \, b_c \, \tau^{-\phi} = b_c \, \Gamma(1-\phi) \, r^{\phi} \, ,
\label{eq:2D_avg_crit}
\end{equation}
where $\Gamma(z)$ is the gamma function.

We have performed Monte Carlo simulations with different values of $r$ to verify the dependence of $\overline{m(r)}$ on $r$. The agreement is excellent in all the three phases (Fig. 4 a), Fig. 4 b) and Fig. 4 c)), especially for small values of $r$. This is consistent with the fact that we have taken the late time evolution of $m(t)$ in Eq.~(\ref{eq:2D_avg_mag}), which holds only for small $r$.

\begin{figure*}[h]
\centering
\includegraphics[width = \linewidth]{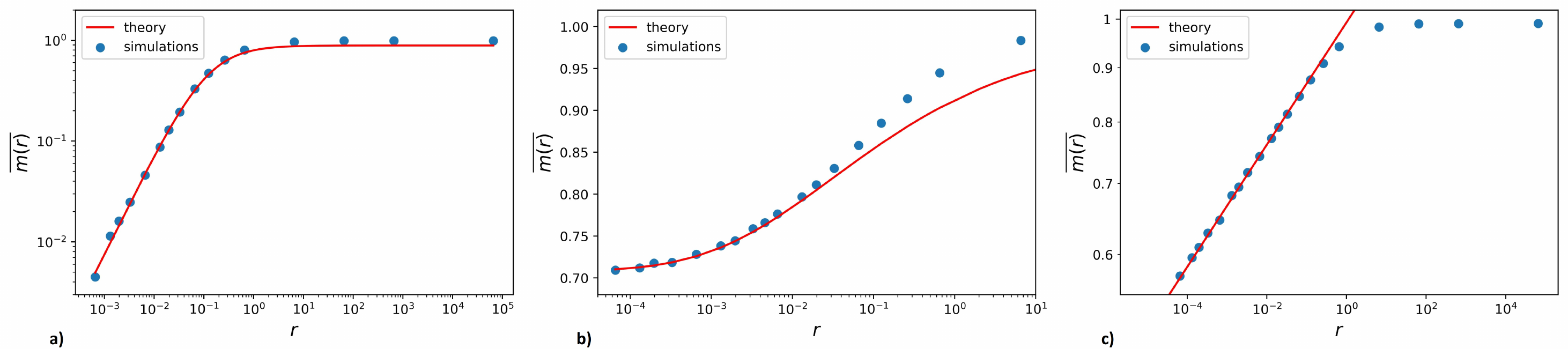}
\caption{Plot of $\overline{m(r)}$ for $T > T_c$, $T<T_c$ and $T=T_c$. The red lines are the theoretical results (Eq.~(\ref{eq:2D_avg_par}), Eq.~(\ref{eq:2D_avg_fer}) and Eq.~(\ref{eq:2D_avg_crit})), while the blue dots come from Monte Carlo simulations.}
\label{fig:totavg}
\end{figure*}

\section{General expression of the cumulative distribution}

In the main text, using renewal theory, we have obtained the expression of the non-equilibrium stationary PDF of the average magnetisation in presence of a resetting rate $r$:
\begin{equation}
P_r^{\text{stat}}(m) = r \int_0^{\infty} d\tau \, e^{-r\tau} \, \delta(m-m(\tau)) \, \;,
\label{eq:trick}
\end{equation}
where $m(\tau)$ denotes the deterministic Glauber evolution of the average magnetisation. It is easy to compute exactly the CDF
\begin{equation}
F_r^{\text{stat}}(m) = \int_{0}^{m} P_r^{\text{stat}}(m') \, dm' \;
\end{equation}
that we plot in Fig. 2 of the main text. We can write $P_r^{\text{stat}}(m)$ as a slightly more general expression
\begin{equation}
P_r^{\text{stat}}(m) = \int_0^{\infty} d\tau \, f(\tau) \, \delta(m-g(\tau)) \; ,
\end{equation}
where of course in our case $f(\tau) = r \, e^{-r\tau}$ and $g(\tau) = m(\tau)$. Then, making the change of variable $z = g(\tau)$, we can write the previous integral as
\begin{equation}
\begin{split}
P_r^{\text{stat}}(m) &=  \int_{g(0)}^{g(\infty)} \frac{1}{g'(g^{-1}(z))} \, f(g^{-1}(z))) \, \delta(m-z) \, dz = \\
&= \frac{1}{|g'(g^{-1}(m))|} \, f(g^{-1}(m))).
\end{split}
\end{equation}
Since $f^{'}(t) = -r \, f(t)$, the corresponding CDF is given by
\begin{equation}
F_r^{\text{stat}}(m) = \pm \, \frac{1}{r} \, f(g^{-1}(m)) + const,
\label{eq:primit}
\end{equation}
where the sign is $+$ and $const = 0$ if $g$ is a decreasing function of time, while the sign is $-$ and $const=1$ if $g$ is an increasing function of time. Since in our case $f(t) = r \, e^{-rt}$, Eq.~(\ref{eq:primit}) reduces to
\begin{equation}
F_r^{\text{stat}}(m) = \pm \, e^{-r(g^{-1}(m))} + const.
\label{eq:cumul}
\end{equation}
This is the CDF that we plot in the main text, with the use of different functions $g$ in the three temperature regimes. The derivation of Eq.~(\ref{eq:cumul}) is based on two properties of the functions $f$ and $g$. The first one is that $f$ and its time derivative $f^{'}$ are proportional, since $f$ is an exponential function. The second property is that the function $g$, which describes the (deterministic) time evolution of the average magnetisation, is a monotonic function and, therefore, is invertible. In particular, this property of the function $g$ is valid in the time window we considered for all the three different temperature regimes (paramagnetic and ferromagnetic phases and at the critical temperature).

\section{Numerical estimate of $r^*(T)$ in the paramagnetic phase of the 2$d$ Ising model}

The exact expression of the crossover line $r^*(T)$ in the 1$d$ Ising model is given by Eq. (7) of the main text as $r^*(T) = 1 - \tanh(\frac{2J}{k_B T})$. It is then interesting to know if a similar curve can be obtained, at least numerically, also in the two dimensional case. For several temperatures $T > T_c$, we perform numerical simulations of the 2$d$ Glauber dynamics in presence of several resetting rates $r$ and detect the value $r^*(T)$ that makes the resulting PDF $P_{r^*}^{\rm stat}(m)$ more similar to the uniform distribution in $[0,m_0]$. We thus expect to obtain a curve $r^*(T)$ that divides the para phase from the new ``pseudo-ferro'' phase as depicted in Fig. 1 of the main text.  

If the time evolution of $m(t)$ was given by a simple exponential decay to 0, then we would find that $r^*(T) = \lambda_1(T)$, as in the one dimensional case. However, this is not the case, since this time dependence holds only at late times. Indeed, the time evolution of the average magnetisation in the paramagnetic phase is given by
\begin{equation}
m(t) = \sum_{i=1}^N a_i e^{-\lambda_i t}
\label{eq:true_evo}
\end{equation}
where $\lambda_1 < \lambda_2 < \dots < \lambda_N$ constitute the whole relaxation spectrum, $a_i$ are coefficients and $N = 2^{L^2}$ is the total number of spin configurations in a $L \times L$ lattice. Eq.~(\ref{eq:true_evo}) is valid \textit{for any time t} and is therefore a generalisation of the simple exponential decay. In the long-time limit the two time evolutions coincide, because only the minimum value among $\lambda_i$'s (i.e. $\lambda_1$) survives as $t \rightarrow \infty$. 

In principle, $N$ should be equal to the total number of spin configurations. Of course, to make a reasonable fit, this number is too big: therefore we choose to take $N = 3$, that is we approximate $m(t)$ to be the sum of the three exponentials with the smallest decay rates. As a consequence, from the fit of $m(t)$ we estimate six parameters, which are $a_i$ and $\lambda_i$ for $i = 1, 2, 3$.

\begin{figure}[h]
\centering
\includegraphics[width=0.45\linewidth]{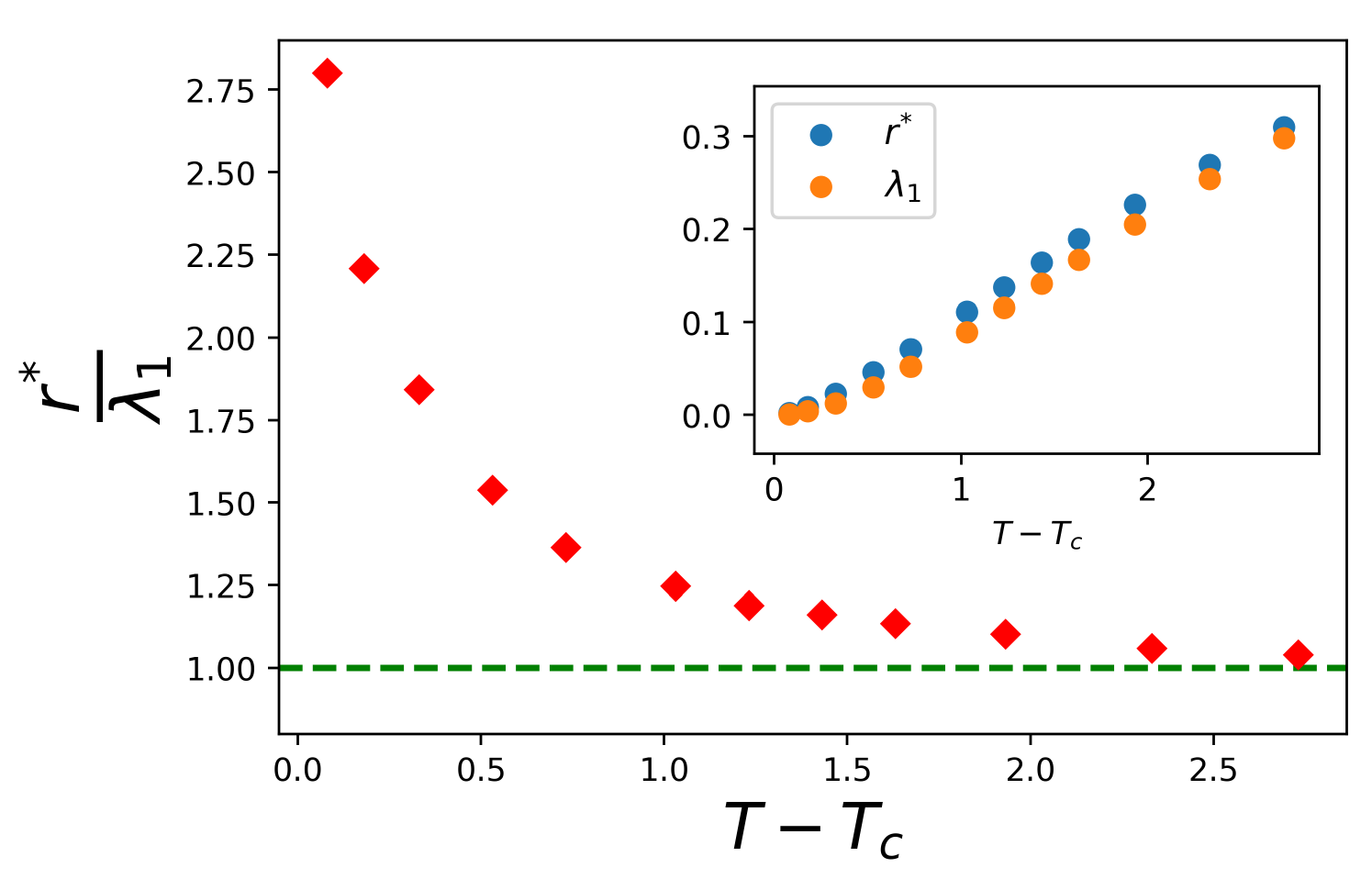}
\caption{The ratio $\frac{r^*}{\lambda_1}$ and, in the inset, the values of $\lambda_1$ and $r^*$ are estimated with Monte Carlo simulations at different temperatures above $T_c$.
\label{fig:definitive}}
\end{figure}

It is therefore interesting to compare the curve $r^*(T)$ obtained as illustrated above with the value $\lambda_1(T)$ obtained by fitting this $m(t)$ to numerical data, for different temperatures $T>T_c$. The inset of Fig.~\ref{fig:definitive} shows the values of $r^*$ and $\lambda_{1}$ for different temperatures above $T_c$. Note that $r^*(T) > \lambda_1(T)$ $\forall T >T_c$. In particular, if we imagine to draw a line connecting all the blue dots, we obtain the red dashed line in the phase diagram in Fig. 1 of the main text. In the main plot of Fig.~\ref{fig:definitive} the ratio $r^*/\lambda_1$ is plotted for different temperatures above $T_c$, using the same data points of the inset. We see that this ratio converges to 1 as the temperature $T$ increases, proving that, when $T \gg T_c$, the approximation $r^* = \lambda_1$ (which is the exact relation found in the $1d$ case) is correct.

